\documentclass[useAMS,usenatbib,psfig]{mn2e}
\usepackage{graphicx}
\begin{document}
\title[Three 2013 novae]{Study of three 2013 novae: V1830 Aql, V556 Ser and V809 Cep}

\author[U. Munari et al.]{U. Munari$^{1}$, P. Ochner$^{1}$, S. Dallaporta$^{2}$, P. Valisa$^{2}$, M. Graziani$^{2}$, G.L. Righetti,$^{2}$
\newauthor G. Cherini$^{2}$, F. Castellani$^{2}$, G. Cetrulo$^{2}$, A. Englaro$^{2}$\\
$^{1}$INAF Astronomical Observatory of Padova, 36012 Asiago (VI), Italy\\
$^{2}$ANS Collaboration, c/o Osservatorio Astronomico, via dell'Osservatorio 8, 36012 Asiago (VI), Italy}

\date{Accepted .... Received ....; in original form ....}

\pagerange{\pageref{firstpage}--\pageref{lastpage}} \pubyear{2010}

\maketitle

\label{firstpage}

\begin{abstract}
$B$$V$$R_{\rm C}$$I_{\rm C}$ photometry and low-, medium- and
high-resolution Echelle fluxed spectroscopy is presented and discussed for
three faint, heavily reddened novae of the FeII-type which erupted in 2013. 
V1830 Nova Aql 2013 reached a peak $V$=15.2 mag on 2013 Oct 30.3 UT and
suffered from a huge $E_{B-V}$$\sim$2.6 mag reddening.  After a rapid
decline, when the nova was $\Delta V$=1.7 mag below maximum, it entered a
flat plateau where it remained for a month until Solar conjunction prevented
further observations.  Similar values were observed for V556 Nova Ser 2013,
that peaked near $R_{\rm C}$$\sim$12.3 around 2013 Nov 25 and soon went lost
in the glare of sunset sky.  A lot more observations were obtained for V809
Nova Cep 2013, that peaked at $V$=11.18 on 2013 Feb 3.6.  The reddening is
$E_{B-V}$$\sim$1.7 and the nova is located within or immediately behind the
spiral Outer Arm, at a distance of $\sim$6.5 kpc as constrained by the
velocity of interstellar atomic lines and the rate of decline from maximum. 
While passing at $t_3$, the nova begun to form a thick dust layer that
caused a peak extinction of $\Delta V$$>$5 mag, and took 125 days to
completely dissolve.  The dust extinction turned from neutral to selective
around 6000 \AA.  Monitoring the time evolution of the integrated flux of
emission lines allowed to constrain the region of dust formation in the
ejecta to be above the region of formation of OI 7774 \AA\ and below that of
CaII triplet.  Along the decline from maximum and before the dust
obscuration, the emission line profiles of Nova Cep 2013 developed a narrow
component (FWHM=210 km/sec) superimposed onto the much larger normal
profile, making it a member of the so far exclusive but growing club of
novae displaying this peculiar feature.  Constrains based on the optical
thickness of the innermost part of the ejecta and on the radiated flux,
place the origin of the narrow feature within highly structured internal
ejecta and well away from the central binary.
\end{abstract}
\begin{keywords}
Stars: novae 
\end{keywords}

\section{Introduction}

Several Galactic novae are regularly missed because of yearly Sun
conjunction with the Galactic central regions, where most of them appear, or
because the heavy interstellar absorption low on the Galactic plane dims
them below the observability threshold.  Others are so fast that remains
above the detection threshold of equipment used by amateur astronomers (who
discover the near totality of Galactic novae), for too short a time to get a
fair chance to be discovered (Warner 1989, 2008; Munari 2012).  Finally, a
significant fraction of those discovered and catalogued lack published data
necessary to properly document and characterized them (Duerbeck 1988).

All these deficiencies reflect into still debated statistics about the basic
properties of the Galactic novae, like their average number per year or
their fractional partnership to Galactic populations like the Bulge, and the
Thin Disk and the Thick Disk (della Valle and Livio 1998; della Valle 2002;
Shafter 2002, 2008).  The latter has long reaching implications about the origin,
birth-rate and evolution of binary systems leading to nova eruption, the
amount and type of nuclearly-processed material returned to the interstellar
medium, the viability of recurrent novae as possible precursors of type Ia
supernovae.

Three heavily reddened and faint novae appeared in 2013. V1830~Aql and
V556~Ser were rapidly lost in the glare of sunset sky, and the spectroscopic
observations here presented could possibly be the only multi-epoch
available.  V809~Cep was more favourably placed on the sky, but its peak
brightnes of just $V$=11.18, the fast decline and the very thick dust cocoon
it rapidly developped one month into the decline required a highly motivated
effort to conduct a thorough monitoring at optical wavelengths.

As a result of our effort to contribute to the documentation of as many as
possible of the less observed novae, in this paper we present the results
and analysis of our $B$$V$$R_{\rm C}$$I_{\rm C}$ photometry and low-,
medium- and high-resolution fluxed spectroscopy of these three novae.

\section{Observations}

$B$$V$$R_{\rm C}$$I_{\rm C}$ photometry of the program novae was obtained
with the Asiago 67/92cm Schmidt camera and various telescopes operated by
ANS Collaboration (N.  11, 30, 62, 73, 157).  Technical details of this
network of telescopes running since 2005, their operational procedures and
sample results are presented by Munari et al.  (2012).  Detailed analysis of
the photometric procedures and performances, and measurements of the actual
transmission profiles for all the photometric filter sets in use with ANS
Collaboration telescopes is presented by Munari and Moretti (2012).  All
measurements on the program novae were carried out with aperture photometry,
the long focal length of the telescopes and the absence of nearby
contaminating stars not requiring to revert to PSF-fitting.  All photometric
measurements were carefully tied to a local $B$$V$$R_{\rm C}$$I_{\rm C}$
photometric sequence extracted from the APASS survey (Henden et al.  2014)
and ported to the Landolt (2009) system of equatorial standards following
the transformation equations of Munari (2012).  The adopted local photometric
sequences were selected to densely cover a color range much larger than that
displayed by the nova during its evolution.  The sequences were intensively
tested during the whole observing campaign for linearity of color equations
and for absence of intrinsic variability of any of their constituent stars. 
The use of the same photometric comparison sequences for all the involved
telescopes and for all observations of the novae, ensues the highest
internal homogeneity of the collected data.  The median value of the total
error budget (defined as the quadratic sum of the Poissonian error on the
nova and the formal error of the transformation from the local to the
standard system as defined by the local photometric sequence) of the
photometric data reported in this paper is 0.010 mag for $B$, 0.008 in $V$,
0.007 in $R_{\rm C}$, 0.007 in $I_{\rm C}$, and 0.008 mag for $B-V$, 0.008
in $V-R_{\rm C}$, and 0.009 in $V-I_{\rm C}$.  Colors and magnitudes are
obtained separately during the reduction process, and are not derived one
from the other.

  \begin{table}
    \centering      
     \caption{Journal of spectroscopic observations of the three program
     novae. $\Delta t$ is counted from optical maximum, except for V556 Ser
     for which the reference epoch is the time of nova discovery. Dispersion
     (\AA/pix) is given for single dispersion spectra, while resolving power
     is given for Echelle multi-order spectra.}
    \includegraphics[width=8.5cm]{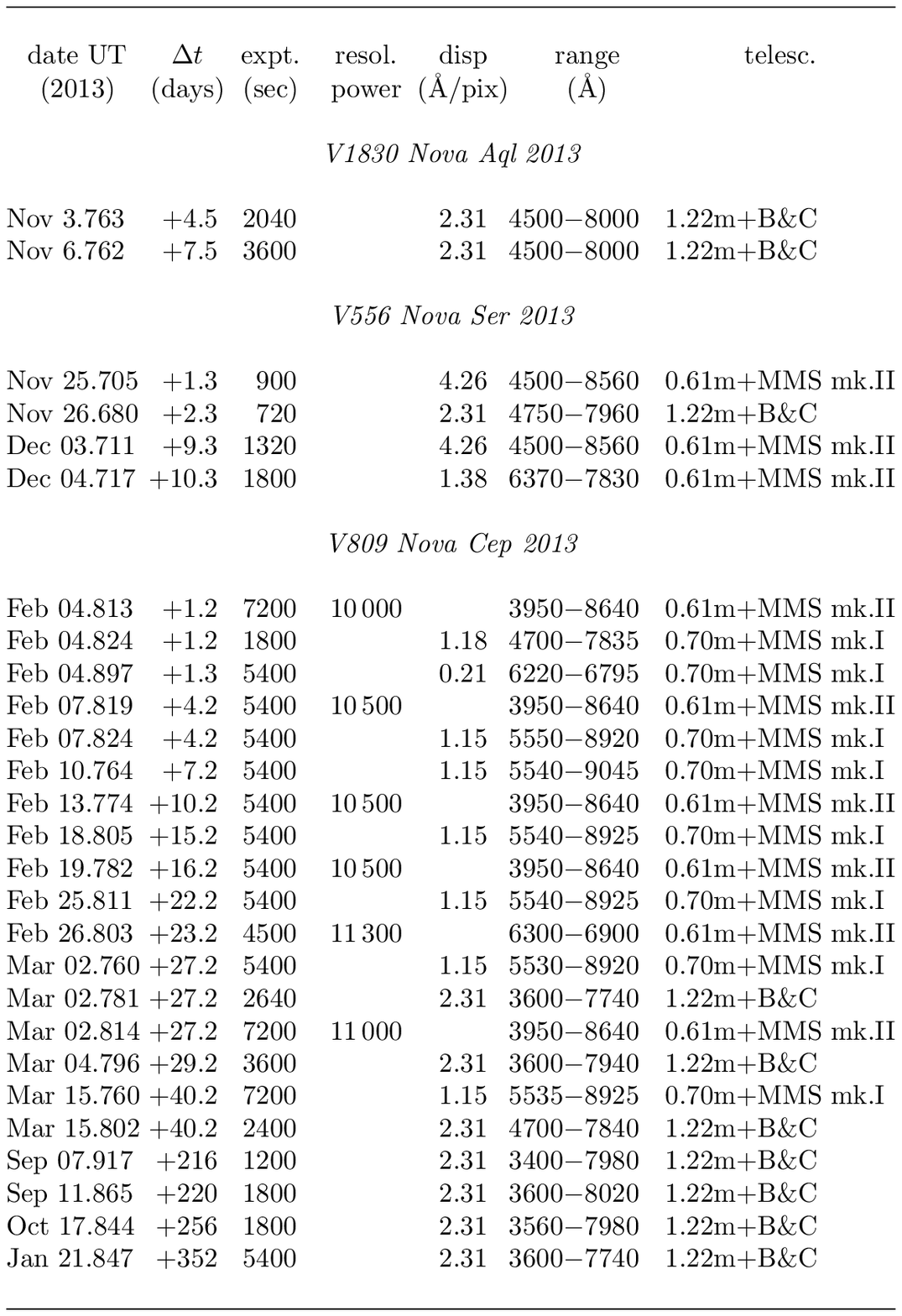}
     \label{tab1}
  \end{table}    

Spectroscopic observations of the program novae
(see Table~1 for a log), have been obtained with three telescopes.

The ANS Collaboration 0.70m telescope located in Polse di Cougnes and
operated by GAPC Foundation is equipped with a mark.I Multi Mode
Spectrograph, and obtain low- and medium-resolution spectra well into the
deep red because its thick, front illuminated CCD (Apogee ALTA U9000,
3056$\times$3056 array, 12$\mu$m pixel, KAF9000 sensor) does not suffer from
noticeable fringing. The ANS Collaboration 0.61m telescope operated by
Schiaparelli Observatory in Varese and equipped with a mark.II Multi Mode
Spectrograph obtains low-, medium- and high-resolution Echelle spectra. The
detector is a SBIG ST10XME with KAF-3200ME chip, 2192$\times$1472 array and
6.8 $\mu$m pixel, with micro-lenses to boost the quantum efficiency. It also
does not suffer from fringing because it is of the thick, front illuminated
type. The optical and mechanical design, operation and performances of
mark.I, II and III Multi Mode Spectrographs in use within ANS Collaboration
are described in detail by Munari and Valisa (2014).

Low resolution spectroscopy of the program novae was obtained also with the
1.22m telescope + B\&C spectrograph operated in Asiago by the Department of
Physics and Astronomy of the University of Padova. The CCD camera is a ANDOR
iDus DU440A with a back-illuminated E2V 42-10 sensor, 2048$\times$512 array
of 13.5 $\mu$m pixels. This instrument was mainly used for the observations
of the faintest states of the program novae. It is highly efficient in the
blue down to the atmospheric cut-off around 3200 \AA, and it is normally not
used longword of 8000 \AA\ for the fringing affecting the sensor.
 
The spectroscopic observations at all three telescopes were obtained in
long-slit mode, with the slit rotated to the parallactic angle, and red
filters inserted in the optical path to suppress the grating second order
where appropriate.  All observations have been flux calibrated, and the same
spectrophotometric standards have been adopted at all telescopes.  All data
have been similarly reduced within IRAF, carefully involving all steps
connected with correction for bias, dark and flat, sky subtraction,
wavelength and flux calibration.

\section{V1830 Aql (Nova Aql 2013)}

Nova Aql 2013 was discovered by K. Itakagi at unfiltered 13.8 mag on Oct. 
28.443 UT (cf.  CBET 3691) at equatorial coordinates
$\alpha$=19$^h$02$^m$33$^{s}$.35 and
$\delta$=$+$03$^\circ$15'19".0, corresponding to Galactic
coordinates $l$=037.12, $b$=$-$01.07.  The variable was designated PNV
J19023335+0315190 when it was posted at the Central Bureau for Astronomical
Telegrams TOCP webpage.  Spectroscopic confirmations as a classical nova of
the {\it FeII}-type were provided by Fujii (2013), Munari (2013) and Takaki
et al.  (2013).  It was later assigned the permanent GCVS designation
V1830 Aql (Samus 2013b).  At the reported astrometric position, no candidate
progenitor is visible on any digitized Palomar I and II Sky Survey plates,
and no counterpart is present in the 2MASS catalog.  In addition to the
discovery observations summarized in CBET 3691, no further information has
been so far published on Nova Aql 2013.

  \begin{figure}
    \centering   
    \includegraphics[width=8.5cm]{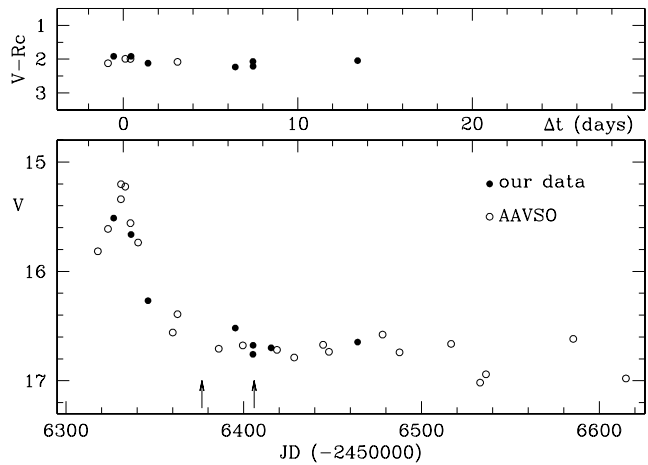}
     \caption{Light- and color-curves of V1830 Nova Aql 2013. $\Delta t$ time 
     is counted from maximum brightness, attained on 2013 Oct 10.3 UT. The
     arrows mark the time of the spectroscopic observations listed in
     Table~1 and shown in Figure~2.} 
     \label{fig1}
  \end{figure}

\subsection{Photometric evolution and reddening}

Our $V$,$R_{\rm C}$ photometry of Nova Aql 2013 is presented in Table~2 and
plotted in Figure~1 together with data retrieved from the public AAVSO
database.  For some dates, the AAVSO data are in the format of a series of
rapid measurements (protracted for hours) from the same observer, that we
have averaged into a single data point in Figure~1.  The reason for that is
the large scatter present in these AAVSO time-series.  The scatter does not
seem related to intrinsic variability of the nova but more likely to the low
S/N of individual data points (short exposures with very small telescopes on
a very faint nova low on the horizon) and to the absence of proper
color-correction of the observations to transfer them from the local to the
standard system (the AAVSO measurements are differential with respect to
radomly choosen nearby field stars).  The very red colors of the nova (far
redder than typical surrounding field stars) and the large changes in
airmass and sky transparencies encountered during these long time-series
runs, suggest the apparent variability present in the AAVSO data to be an
observational artifact.

  \begin{figure*}
    \centering   
    \includegraphics[width=16.5cm]{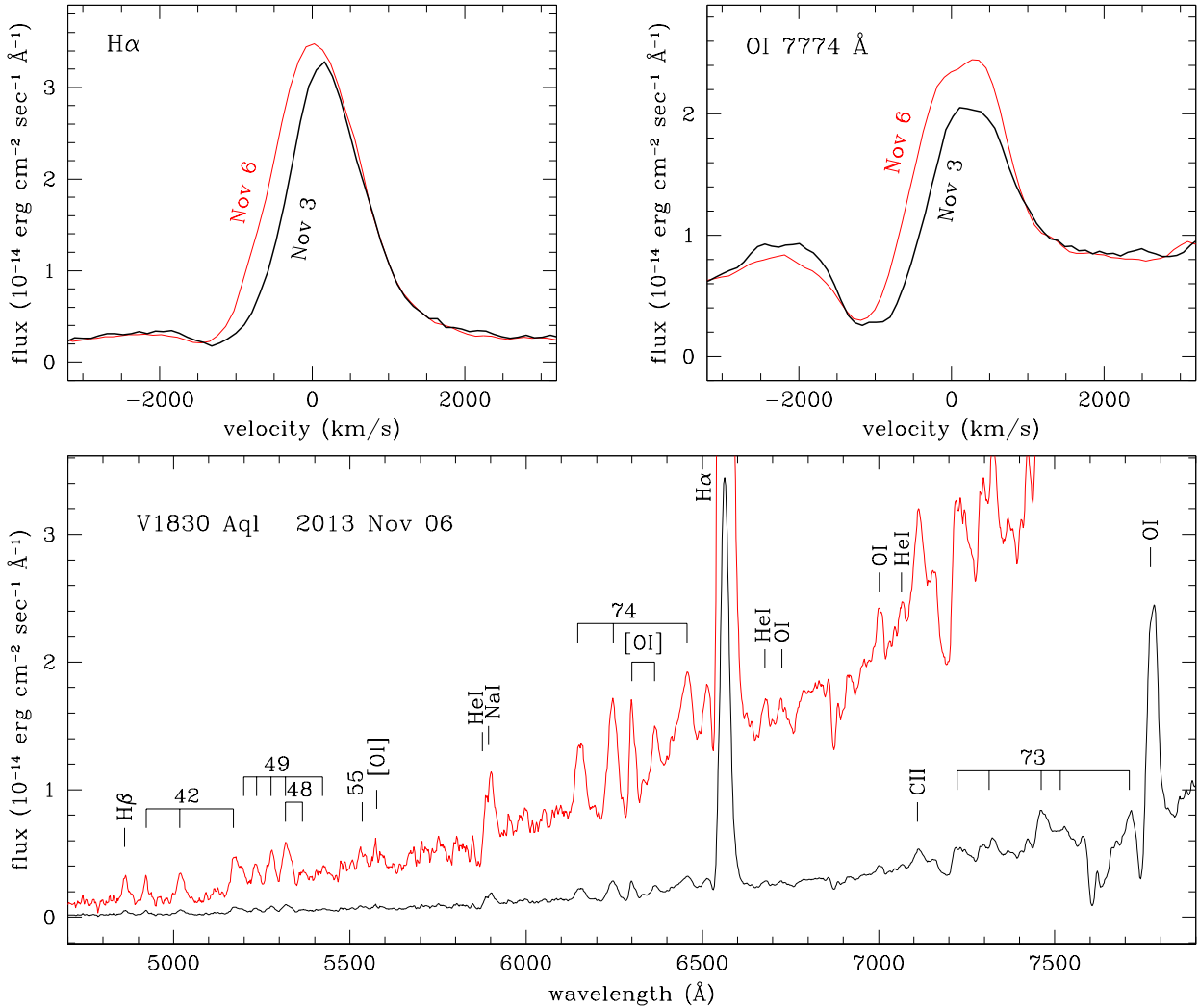}
     \caption{Spectroscopy of V1830 Nova Aql 2013. {\it Bottom}: spectrum
     for 6 November 2013 with strongest emission lines identified. The
     numbers mark FeII multiplets. The spectrum is plotted twice, at full
     flux scale and 6$\times$ to emphasize visibility of weak features and
     continuum slope.
     {\it Top}: comparison of the H$\alpha$ and OI 7774 \AA\ lines profiles
     for the two observing dates, 2013 November 3 and 6.} 
     \label{fig1}
  \end{figure*}  

  \begin{table}
    \centering      
     \caption{$V$$R_{\rm C}$ photometry of V1830 Nova Aql 2013.}
    \includegraphics[width=6.5cm]{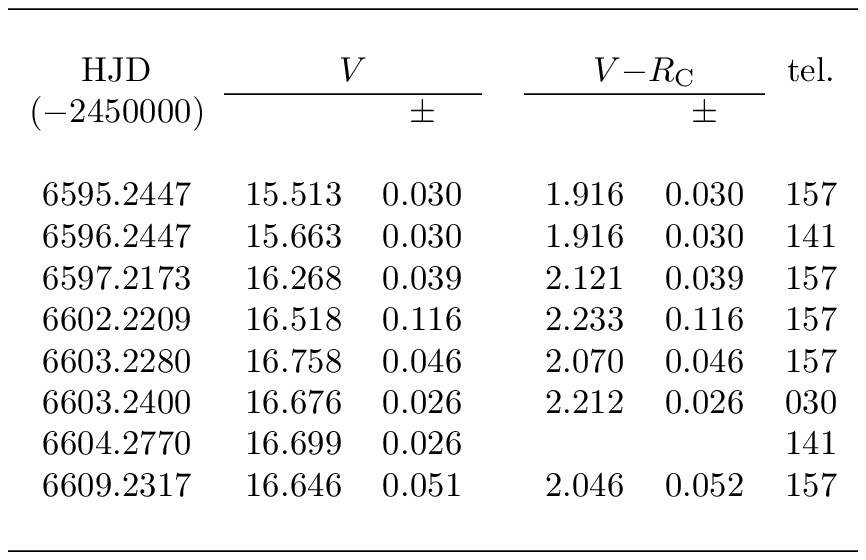}
     \label{tab2}
  \end{table}    

  \begin{table}
    \centering      
     \caption{Integrated emission line fluxes (in units of 10$^{-15}$ erg
     cm$^{-2}$ sec$^{-2}$) measured on the 6 Nov 2013 spectrum of V1830 Nova 
     Aql 2013 shown in Figure~2.}
    \includegraphics[width=6.7cm]{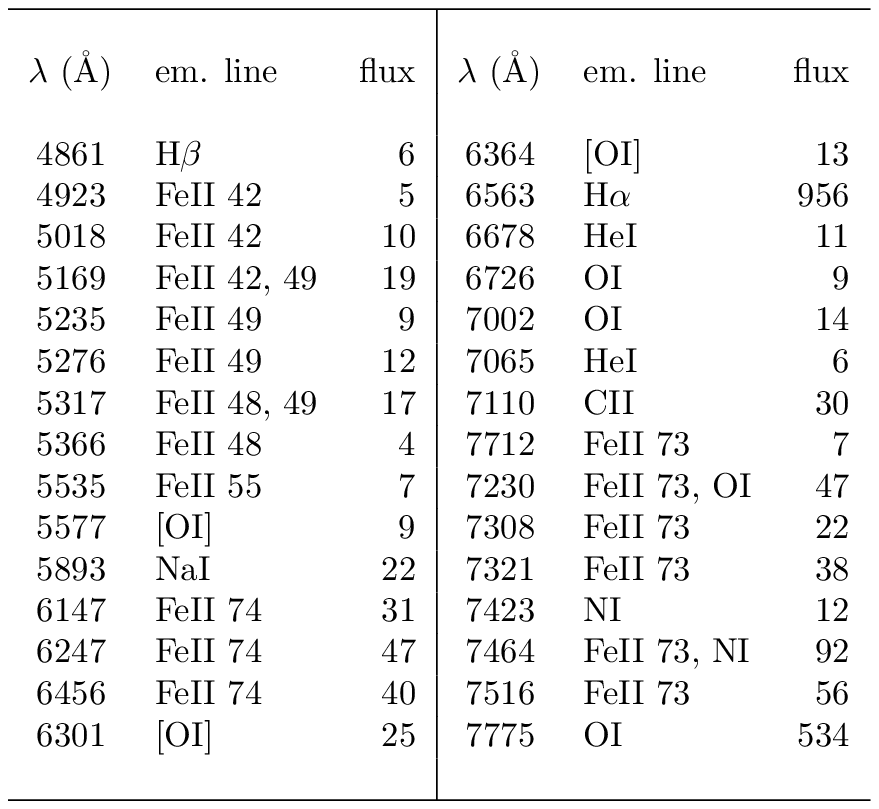}
     \label{tab3}
  \end{table}

The lightcurve in Figure~1 indicates that the nova was discovered during its
rapid rise to maximum.  The final portion of the rise to maximum shown in
Figure~1 was covered at a rate of 0.45 mag day$^{-1}$.  The maximum was
reached at $V_{\rm max}$=15.2$\pm$0.05 on HJD$_{\rm max}$=2456595.8$\pm$0.2
($\equiv$ Oct 30.3 UT), that will be taken as reference $t_\circ$ in the
following.  The $V$$-$$R_{\rm C}$ color has remained constant around 2.0
during the recorded photometric evolution, suggesting a very high reddening
affecting Nova Aql 2013.  In fact, the high galactic latitude and low
reddening ($E_{B-V}$=0.18) FeII-type V339 Nova Del 2013 displayed at maximum
brightness a color $V$$-$$R_{\rm C}$=+0.085 (Munari et al.  2013a).  Supposing the
reddening corrected ($V$$-$$R_{\rm C}$)$_\circ$=$-$0.03 of Nova Del 2013
applies also to Nova Aql 2013 at the time of its maximum brightness, the
reddening corresponding to the observed $V$$-$$R_{\rm C}$=+2.0 would be
$E_{B-V}$$\sim$2.6 (following the reddening analysis of Fiorucci and Munari
2003 for the Johnson-Cousins photometric system as realized by the Landolt
equatorial standards onto which our photometry is accurately placed).

Upon reaching maximum brightness, Nova Aql 2013 bounced off it and
immediately begun to decline, as Figure~1 shows. The early decline from
maximum was very fast: the first $\Delta V$=1.2 mag were covered in just 1.6
days, at a rate of 0.75 mag day$^{-1}$. The decline than quickly leveled off
and proceeded at a much lower pace, about 0.015 mag day$^{-1}$ for several
weeks. The monitoring of the photometric evolution of the nova did not
extend after the first month as it became too faint and progressively lost
in the bright sky at sunset because of the approaching conjunction with the
Sun.

Overall, the light-curve of Nova Aql 2013 in Figure~1 can be best described
as a month-long flat plateau following an initial brief surge that
brightened the nova about $\Delta V$=1.7 mag above the plateau level.  As
such, the application of standard MMRD (absolute Magnitude at Maximum vs. 
Rate of Decline) relations would be doubtful.  Moreover, at the time the
observations had to be terminated because of the conjunction with the Sun,
the nova had not yet declined 2 mag from maximum (cf.  Figure~1), the
minimum amount for which MMRD relations are calibrated.

\subsection{Spectroscopy}

Our fluxed Nov. 6 spectrum of Nova Aql 2013 is shown in Figure~2, its S/N at
blue wavelengths being appreciably higher that the spectrum for Nov. 3. The
high reddening is evident in the continuum steeply rising toward the red.
The spectrum is typical of a FeII-type nova soon after maximum brightness.
The usual FeII multiplets N. 42, 48, 49, 55, 73, 74 are readily seen in
strong emission (compare with the intensity of nearby H$\beta$ in Figure~2),
with additional emissions mainly from NaI and OI. The integrated flux of the
measurable emission lines is listed in Table~3, where the impressive flux
ratio H$\alpha$/H$\beta$=160 stands out, another indication of the huge
reddening affecting this nova.

  \begin{figure*}[!Ht]
    \centering   
    \includegraphics[width=16.5cm]{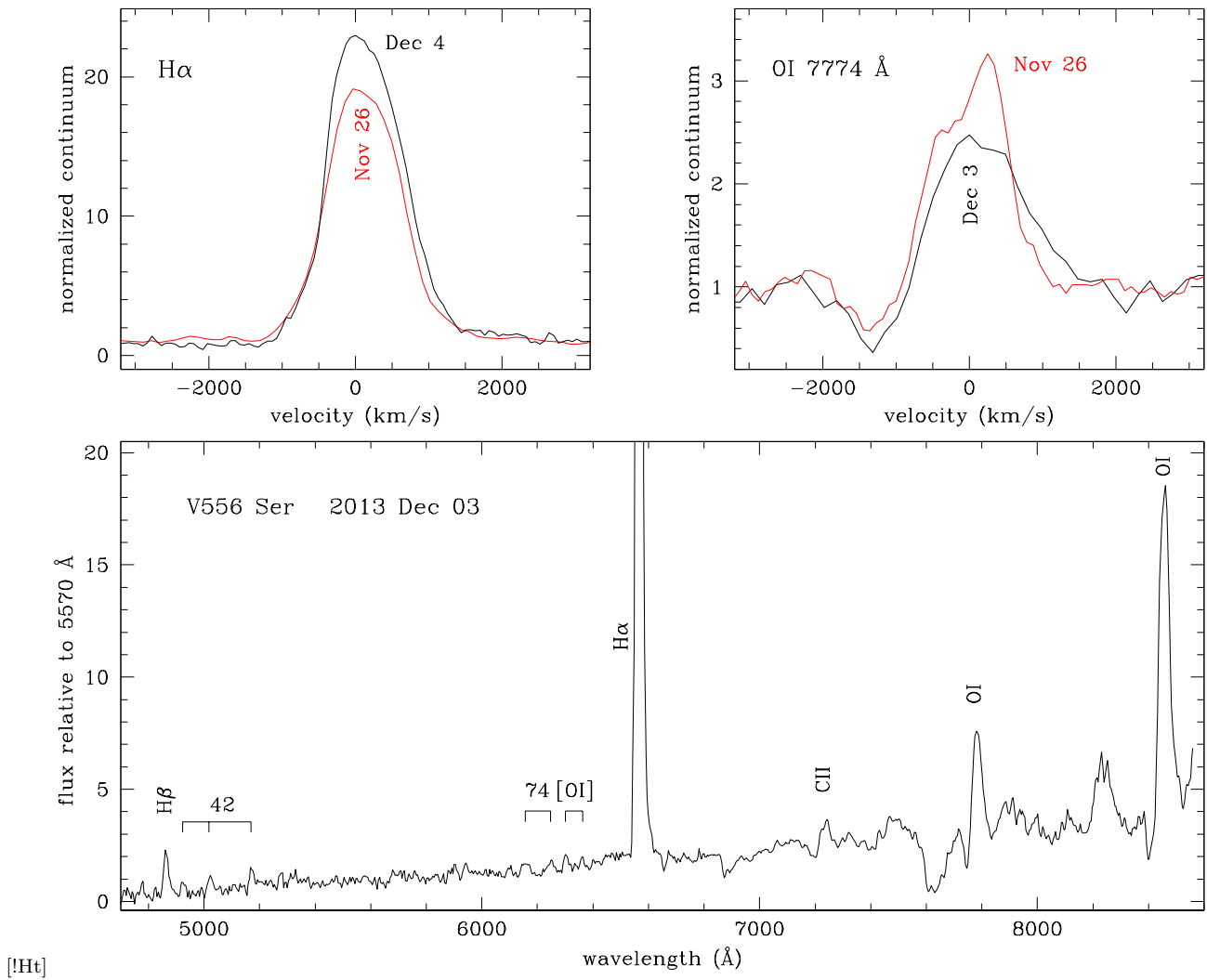}
     \caption{Spectroscopy of V556 Nova Ser 2013. {\it Bottom}: spectrum for
     3 December 2013 with strongest emission lines identified. The numbers
     mark FeII multiplets.  {\it Top}: comparison of the H$\alpha$ and OI
     7774 \AA\ lines profiles for the indicated observing dates.}
     \label{fig1}
  \end{figure*}

The P-Cyg line profiles of H$\alpha$ and OI 7774 \AA\ for the two observing
dates are highlighted and compared in Figure~2 (top panels).  The width of
the emission component of both profiles increased between the two dates:
from 1090 to 1435 km/sec for H$\alpha$, and from 990 to 1150 km/sec for OI. 
The blue shift of the absorption component increased between the two dates:
from $-$1350 to $-$1500 km/sec for H$\alpha$, and from $-$1090 to $-$1200
km/sec for OI 7774 \AA.  This behavior can be interpreted in the simplified
framework of ballistic launched spherical ejecta.  As ionization spreads
through a widening layer of the ejecta, the neutral part is confined to
outer, faster moving regions (increasing the negative velocity of the P-Cyg
absorption components), while reduction of the optical thickness allows
radiation from different parts of the ejecta to contribute to widen the
emission component.  The H$\alpha$ profiles can be accurately fitted by one
Gaussian in emission and another in absorption.  Some structure is instead
present in the OI lines, where the absorption component for Nov 3 is wide
and flat bottomed but otherwise symmetrical, while the profile for Nov 6
shows a flat-topped emission component and a sharper absorption with an
extended blue wing, suggesting absorbing material radially distributed over
a large interval characterized by a range of outwardly increasing
ballistic velocities and decreasing local densities.

\section{V556 Ser (Nova Ser 2013)}

Nova Ser 2013 was discovered as an optical transient by K. Itakagi at
unfiltered 12.3 mag on Nov 24.384 UT (cf.  CBET 3724) at equatorial
coordinates $\alpha$=18$^h$09$^m$03$^{s}$.46 and
$\delta$=$-$11$^\circ$12'34".5, corresponding to Galactic
coordinates $l$=18.21 $b$=+04.17.  The variable was designated PNV
J18090346-1112345 when it was posted at the Central Bureau for Astronomical
Telegrams TOCP webpage.  Spectroscopic confirmation and classification as a
nova was provided by Munari and Valisa (2013) and Itoh et al.  (2013). It
was assigned the permanent GCVS designation V556 Ser (Kazarovets 2013). 
In addition to the discovery observations summarized in CBET 3724, no
further information has been so far published on Nova Ser 2013.  At the reported
astrometric position, no candidate progenitor is visible on digitized
Palomar I Sky Survey and SERC plates, and no counterpart is present in the
2MASS catalog.  The amplitude of the outburst is therefore at least 9 mag in
the optical, and the companion of the erupting white dwarf does not seem to
be evolved away from the Main Sequence.

When it was discovered, the nova was only briefly observable very low on the
horizon after sunset, when the sky was still bright, making it a very
difficult target for observations.  This is why we were not able to collect
photometric observations of this nova, and why there are no observations
logged in the public databases of any amateur astronomer organizations we
consulted, VSNET and AAVSO included.  The only available photometric data
are those collected at the time of discovery by a few observers, and
summarized in CBET 3724 that announced the nova.  They all come from unfiltered
instrumentation and no information is provided about what photometric
sequence and what photometric band was used for their reduction/measurement. 
The nova was reproted at $>$13 mag on Nov.  22.370 and 23.361, 12.3 mag on
Nov 24.384, 11.7 mag on Nov 26.369, and 12.7 mag on Nov 26.373.  The last
two measurements are essentially simultaneous and still differ by 1 whole
magnitude, testifying the unfiltered nature of these measurements, their
difficulty and their uncertainty.  The best it can be said is that the
magnitude of the nova seems to have remained constant around unfiltered
12.2/12.3 mag during the interval Nov 24-26, which could also mark the time
of maximum.  In Table~1 we adopt the time of discovery on Nov 24.384 as the
reference $t_\circ$.

Our spectroscopic observations of Nova Ser 2013, are summarized in Table~1
and cover an interval of 9 days.  The best exposed spectrum is that of Dec
3, which is displayed in Figure~3, together with details from the spectra of
Nov 26 and Dec 4.  While relative fluxes should be correct for the Dec 3
spectrum, the critical observing condition does not allow to confidently fix
the zero point of the flux scale. Consequently, the flux is expressed
relative to that at 5570 \AA, the effective wavelength for the $V$-band
for a highly reddened nova (Fiorucci and Munari 2003).

The continuum of the Dec 3 spectrum in Figure~3 rises steeply toward the red
with an inclination corresponding to $V$$-$$R_{\rm C}$=+1.62, indicative of
a high reddening.  The strongest emission lines belong to the Balmer series
and OI.  The H$\alpha$/H$\beta$ flux ratio is $\sim$32, confirming the high
reddening.  The OI 8446 / OI 7774 flux ratio is 3.8, reversed with respect
the pure recombination value of 3/5.  Even if the large reddening
contributes in depressing OI 7774 with respect to OI 8446, it alone cannot
justify the large 3.8 flux ratio, which therefore suggests that on Dec 3 the
fluorescent pumping of OI 8446 by hydrogen Ly-$\beta$ (as originally
proposed by Bowen 1947), was already effective.

Weak emission lines originate from FeII multiplets 42 and 74, and possibly
49 and 73, that allows to classify Nova Ser 2013 as a {\it FeII-nova}
following Williams (1992). The [OI] 6300, 6364
\AA\ doublet is faintly visible in emission, with a flux ratio of 2.2, 
indicating mild optically thick conditions.

P-Cyg absorption components are clearly visible in OI 7774 and 8446 \AA\
lines, and absent in Balmer lines (cf.  Figure~3).  They are symmetric
and well fitted by a Gaussian component.  On Nov 25 and 26 spectra the
absorption is blue shifted by 1400 km/s with respect to the emission
component, and by a similar amount on Dec 3 and 4 spectra.  On the Dec 3
spectrum, P-Cyg absorption components to FeII multiplet 42 emission lines
are present with the same blue-shift as seen in OI 7774 and 8446 lines.  The
two spectra with the highest resolving power are those for Nov 26 and Dec 4. 
Their H$\alpha$ profiles are compared in the top-left panel of Figure 3.
Their FWHM is 1145 km/sec for Nov 26 and 1240 km/sec for Dec 4. The profile for Nov 26
closely resemble a perfect Gaussian, that for Dec 4 is non-symmetric, with a
sharper blue side, perhaps suggestive of a superimposed (but not resolved by
our low resolution spectra) absorption component or emerging sub-structures.

  \begin{figure}
    \centering   
    \includegraphics[width=8.5cm]{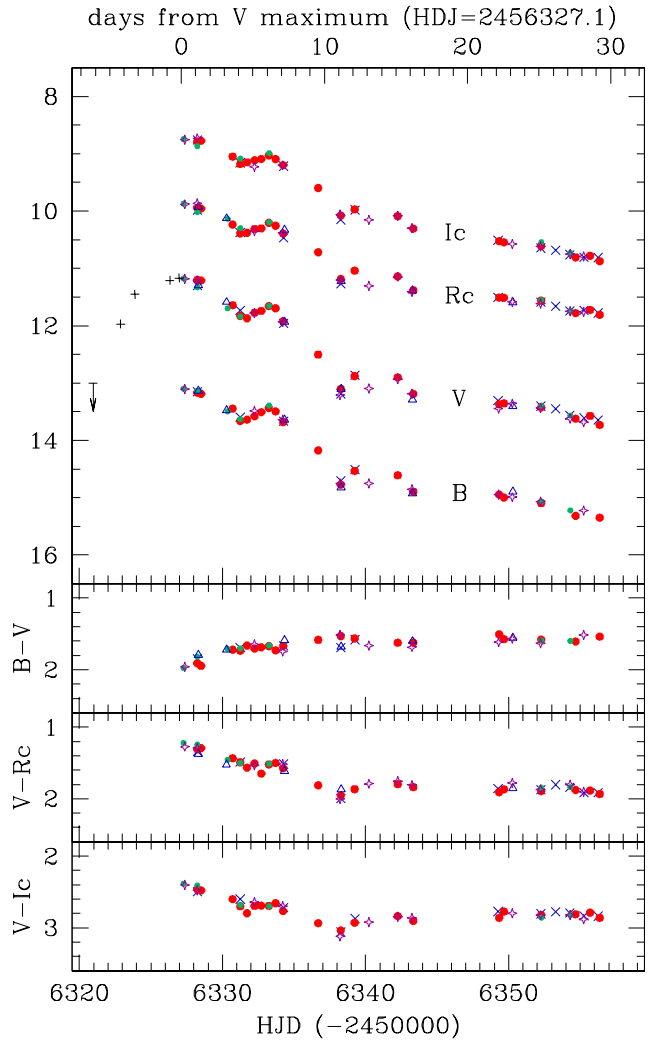}
     \caption{Early photometric evolution of Nova V809 Cep 2013, before the
     dust started to condense in the ejecta.  A legend for the symbols
     identifying different ANS telescopes is provided in Figure 5.  The
     crosses in the early portion of the $V$ band light-curve are
     discovery and pre-discovery data from CBET 3397.}
     \label{fig1}
  \end{figure}

\section{V809 Cep (Nova Cep 2013)}

Nova Cep 2013  was discovered by K. Nishiyama and F.  Kabashima at
unfiltered 10.3 mag on Feb 2.412 UT (cf.  CBET 3397) at equatorial
coordinates $\alpha$=23$^h$08$^m$04$^{s}$.71 and
$\delta$=$+$60$^\circ$46'52".1, corresponding to Galactic
coordinates $l$=110$^{\circ}_{.}$65, $b$=+00$^{\circ}_{.}$40.  It was
designated PNV J23080471+6046521 and spectroscopic classification as a {\em
FeII} nova was provided by Imamura (2013) and by Ayani and Fujii (2013).
It was assigned the permanent GCVS designation V809 Cep (Samus 2013a).  

Munari et al. (2013b) found in mid-March that the nova was rapidly declining
in brightness and inferred it was forming a thick layer of dust in its
ejecta.  This was later confirmed by the infrared observations from Raj et
al.  (2013) and Ninan et al.  (2013) that found a large infrared excess at
$J$$H$$K$ bands and a featureless spectrum at $K$ wavelengths.

Radio observations of Nova Cep 2013 were obtained by Chomiuk et al. (2013). 
They detected the nova at 7.4 and 36.5 GHz (4.0 and 0.82 cm) on Mar 22.5, 47
days past maximum optical brightness and 11 days since the onset of dust
condensation in the ejecta.  Similar and earlier observation performed 10
days past optical maximum yield no detection.  Negative radio detection at
1.3 and 0.61 GHz (23 and 49 cm) on June 1 and Aug 23, respectively, led
Dutta et al.  (2013) to place upper limits to any non-thermal radio emission
from the nova.

Finally, Chomiuk et al. (2013) tried to record X-ray and ultraviolet
emission from Nova Cep 2013 with Swift satellite observations carried out
Feb 8.3 and 22.1 (4.7 and 18.5 days past optical maximum).  The nova was not
detected in X-rays on both epochs, while the UVOT UVM2 ultraviolet
magnitudes (central wavelength = 2246 \AA) were 20.0 and 19.5, respectively.

  \begin{figure}
    \centering   
    \includegraphics[width=8.5cm]{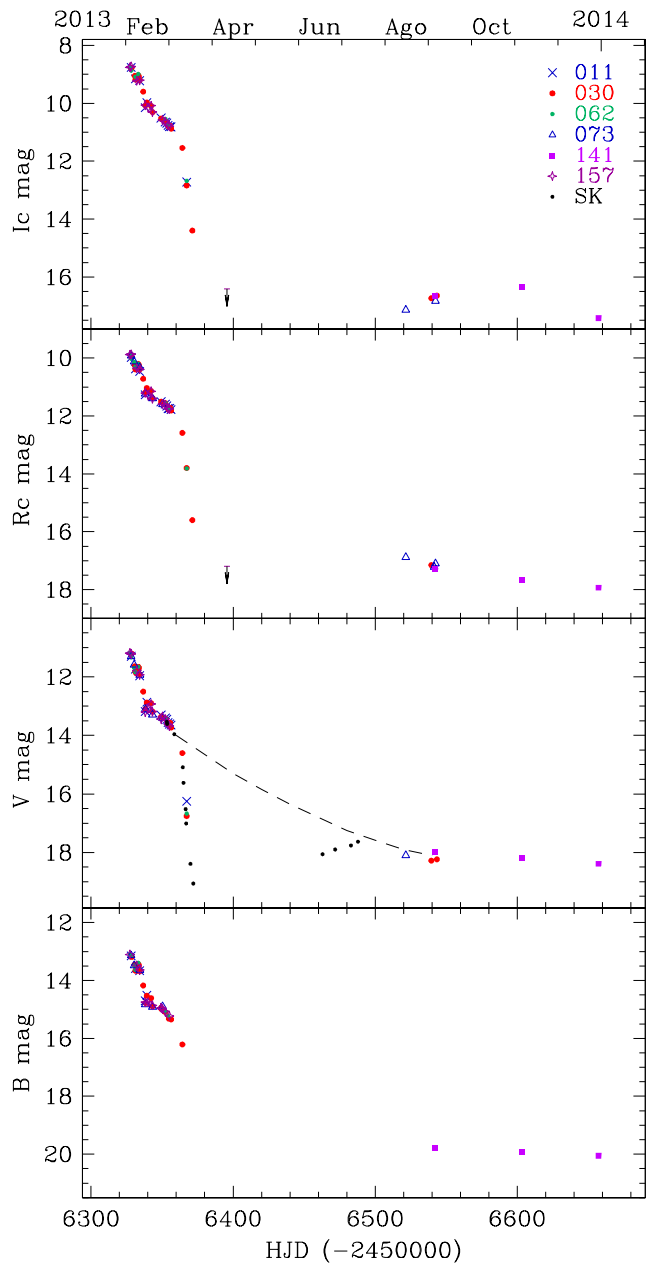}
     \caption{Overall photometric evolution of V809 Nova Cep 2013, showing
     the onset of dust condensation in March 2013, and emergence from it
     around mid-2013. The legend identifies the ANS Collaboration telescopes
     contributing the data to this and Figure~4. SK identifies $V$ band data
     from VSNET observer Seiichiro Kiyota.}
     \label{fig1}
  \end{figure}

\subsection{Photometric evolution}

Our $B$$V$$R_{\rm C}$$I_{\rm C}$ photometric observations of Nova Cep 2013
are presented in Table~4 and plotted in Figures~4 and 5.  Combining with the
pre-discovery data summarized in CBET 3397, it allows to fix the maximum to
have occoured on Feb 3.6$\pm$0.2 (JD=2456327.1) at $V$=11.181 mag,
$B$$-$$V$=+1.967, $V$$-$$R_{\rm C}$=+1.248, $V$$-$$I_{\rm C}$=+2.394.  These
colors are {\em very} red for a nova and suggest a huge interstellar
reddening.  The time required to decline by 2 and 3 mag has been
$t^{V}_{2}$=16, $t^{V}_{3}$=36 days for the $V$ band and $t^{B}_{2}$=25,
$t^{B}_{3}$=37 days for the $B$ band, respectively.  Following the
classification summarized by Warner (1995), they qualify Nova Cep 2013 as a
{\it fast} nova.  The magnitude 15 days past maximum was $t^{V}_{15}$=13.07
and $t^{B}_{15}$=14.75 mag.

Soon after reaching maximum, Nova Cep 2013 started the decline that proceeded
smoothly for the first four days, after which the nova took two days to rise
toward something resembling a secondary maximum, and then resumed the decline but
with some fluctuations up to 0.25 mag in amplitude. Eventually, by day
$\sim$20 and $\sim$2.0 mag below maximum, the nova settled onto a smooth
decline that proceeded until day $\sim$36 when, while the nova was 
passing right through $t_{3}$, it begun to rapidly form dust in the ejecta. 

The nova progressively recovered from the dust obscuration episode during
June and July, and by late July it resumed the normal decline, that
continued smoothly until our last photometric observation, 330 days past the
optical maximum when the nova was measured at $V$=18.39, $\Delta V$=7.2 mag
fainter than maximum.

\subsection{Dust formation }

  \begin{table*}
    \centering      
    \caption{Our $B$$V$$R_{\rm C}$$I_{\rm C}$ of V809 Nova Cep 2013 (the
    table is published in its entirety in the electronic edition of this
    journal.  A portion is shown here for guidance regarding its form and
    content).}
    \includegraphics[width=14cm]{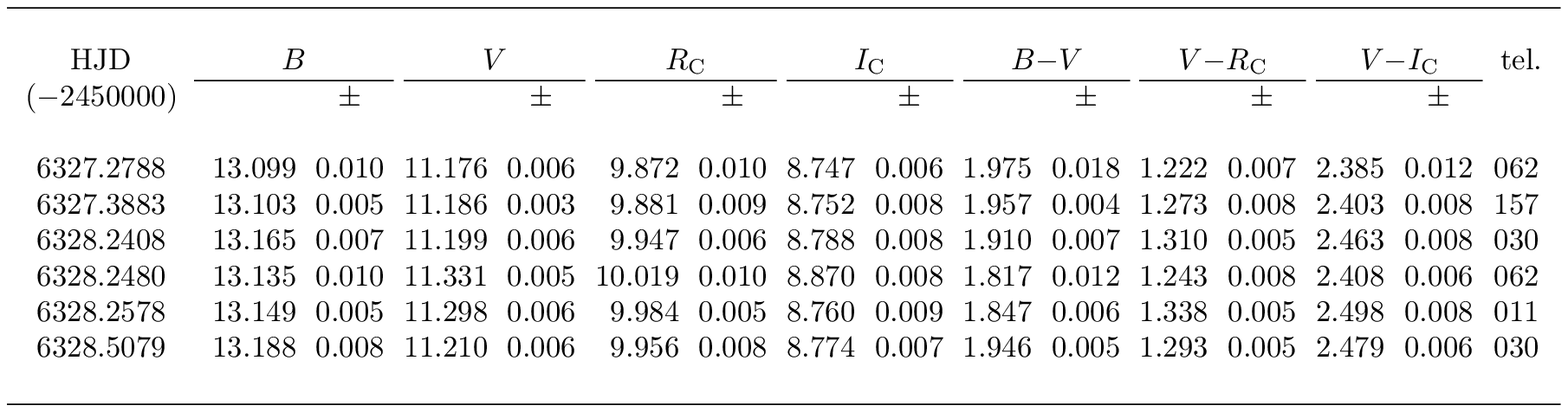}
     \label{tab2}
  \end{table*}

Most dust-forming novae start the condensation when they are between 3 and 4
mag below maximum (McLaughlin 1960, hereafter McL60; Warner 1989).  Assuming
a smooth photometric behavior, the available data allows to fix the onset of
dust formation to March 12.0 $\pm$0.5 UT (JD=2456363.5), when the nova was
at $V$=14.2 and exactly transiting at $t^{V}_{3}$.  Our closest in time
$B$$-$$V$ observation was obtained on March 13.767, when dust condensation
had just begun and was causing a fading of $\Delta V$=0.40 mag.  We measured
$B$$-$$V$=1.49, essentially the same (or even slightly bluer) as before the
onset of dust condensation, while $V$$-$$R_{\rm C}$ and $R_{\rm
C}$$-$$I_{\rm C}$ rapidly became redder.  The fact that the
wavelength-dependent absorption efficiency of the dust turned from neutral
to selective around $\sim$6000~\AA\ suggests a prevalent carbon composition
with a diameter of dust grains of the order of 0.1 $\mu$m (Draine and Lee
1984; Kolotilov, Shenavrin and Yudin 1996).  Nova Cep 2013 behaved closely
similar to Nova Aql 1993 (Munari et al.  1994): this FeII nova was
characterized the same $t_{2}$, $t_{3}$ of Nova Cep 2013, similarly started
to condense dust when transiting exactly at $t^{V}_{3}$, and its dust also
turned from neutral to selective absorption around $\sim$6000~\AA.  It is
also worth noticing that Nova Cep 2013 is placed right on the relation by
Williams et al.  (2013) between the decline time $t_2$ and the time of dust
condensation $t_{\rm cond}$.

The dust condensation in Nova Cep 2013 proceeded at a fast pace: the nova
dropped by $\Delta V$=2.5 during the first 3.0 days, and after 8.5 days it
was 4.7 mag below the extrapolated decline in absence of dust condensation. 
For comparison, the prototype of dust-condensing novae, Nova DQ Her 1934,
dropped by $\Delta m_{\rm vis}$=4.0 mag in 6.6 days (Martin 1989).  The
shape of the light-curve in Figure~5 strongly suggests that dust continued
to condense for a while past JD 2456372 (when the last observation of the declining
branch in Figure 5 was obtained), thus bringing the peak extinction well in
excess of 5 mag in $V$ band (in DQ Her it was $\Delta m_{\rm vis}^{\rm
tot}$=8.0 mag).  The dust layer was therefore completely optically thick at
visual wavelengths.  The fast pace of dust condensation suggests that the
sticking efficiency of condensible elements onto dust grain nucleation sites
was very high.  The nova was re-acquired at optical wavelengths by mid June,
and by mid July the emersion from the dust obscuration was completed and the
nova resumed the normal decline.  From the start of dust condensation to the
end of obscuration, about 125 days passed ($\sim$115 days in DQ Her).

There are a few infrared observations of Nova Cep 2013 obtained at the time
of dust condensation.  On April 26.95, 82 days past maximum brightness and
46 days past the onset of dust condensation, Raj et al.  (2013) measured
Nova Cep 2013 at $J$=13.2, $H$=10.6, $K$=8.2.  The large $J$$-$$K$=+5.0
color corresponds to a black-body temperature of the order of 700~K.  Such
low temperatures are seen in novae forming thick layers (Gehrz et al.  1992,
Gehrz 2008, Evans and Gehrz 2012), while the temperature is hotter in novae with 
thinner dust shells (Mason et al.  1996, Munari et al.  2008, Banerjee and
Ahosk 2012).  In both cases, the temperature of the dust declines with time,
as the ejecta dilute in the circumstellar space.  According to observations
by Ninan et al.  (2013), the infrared brightness of Nova Cep 2013 further
increased after Raj et al.  (2013) performed their observations.  On June
23, Ninan et al.  obtained $K$=7.40, when the nova was quite advanced in the
re-emersion from the dust obscuration, being just $\Delta V$=1.0 mag below
the extrapolated decline in absence of dust (cf Figure~5).  On July 8, at
$\Delta V$=0.5 mag, Ninan et al.  (2013) measured $K$=7.84.  On the same
date, they recorded a spectrum in the wavelength range covered by the $K$
band (from 2.04 to 2.35 $\mu$m) and found a featureless continuum still
dominated by emission from dust.

  \begin{figure*}
    \centering   
    \includegraphics[angle=270,width=17.5cm]{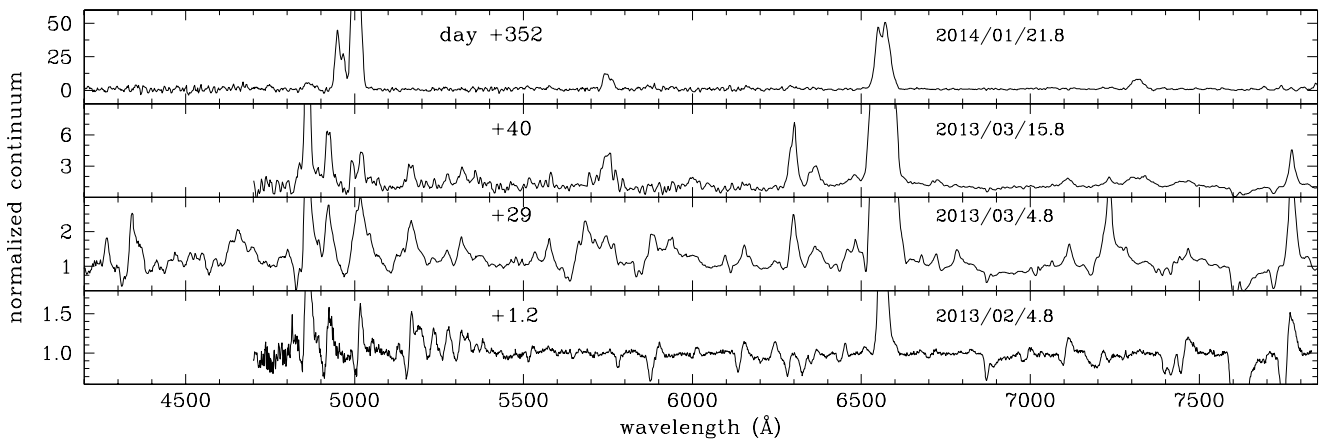}

     \caption{Sample continuum normalized spectra highlighting the overall
              spectral evolution of Nova V809 Cep 2013. Note the different
              ordinate scales. The strongest lines are truncated to emphasize 
              visibility of weak features.}
     \label{fig1}
  \end{figure*}

\subsection{Reddening}

van den Bergh and Younger (1987) derived a mean intrinsic color
$(B-V)_\circ$=$+$0.23 $\pm$0.06 for novae at maximum and
$(B-V)_\circ$=$-$0.02 $\pm$0.04 for novae at $t_2$.  For Nova Cep 2013 we
measured $B$$-$$V$=+1.97 at the time of maximum and $B$$-$$V$=+1.62 at
$t_{2}$ (averaging from $B$$-$$V$=+1.64 at $t^{V}_{2}$, and $B$$-$$V$=+1.60
at $t^{B}_{2}$).  Comparing with van den Bergh and Younger (1987) intrinsic
colors, this indicates a reddening of $E_{B-V}$=1.7.

Munari (2014) has calibrated on many Galactic novae a relation between
$E_{B-V}$ and the equivalent width of the diffuse interstellar band visible
at 6614 \AA.  This DIB has an equivalent width of 0.378 \AA\ on our Echelle
spectrum of Nova Cep 2013 for Feb 07.819, the one with the highest S/N, and
the corresponding reddening is $E_{B-V}$=1.66.

Our Echelle spectra show that both lines of the NaI D$_{1,2}$ doublet are
splinted into two lines of nearly equal intensity, at heliocentric radial
velocities of $-$14 and $-$57 km/s.  Comparing with the Brand and Blitz
(1983) map of the velocity field of the interstellar medium in our Galaxy,
the $-$14 km/s component is associated with the crossing of the Perseus
Arm, while the $-$57 km/s with the crossing the Outer Arm.  The line of
sight to Nova Cep 2013 crosses the Outer Arm at a distance of $\sim$6 kpc
from the Sun, which is therefore a lower limit to the distance of the nova. 
The NaI interstellar lines are core saturated at the high reddening
affecting Nova Cep 2013, while the KI 7699 \AA\ interstellar line is still
on the optically thin linear part of its relation with $E_{B-V}$.  Our
Echelle spectra are somewhat noisy at such red wavelengths, and we have
independently repeated several times the measurement of KI 7699 \AA\ line on
our Echelle spectra for Feb 04.813 and Feb 07.819, finding equivalent widths
between the extrema 0.37 and 0.43 \AA.  Following the calibration by Munari
and Zwitter (1997) this corresponds to 1.6$\leq$$E_{B-V}$$\leq$1.9, in
excellent agreement with the values derived from the color of the nova and
the intensity of the DIB 6614 \AA.

We therefore adopt in this paper $E_{B-V}$=1.7 as the reddening affecting
Nova Cep 2013, that translates into an extinction of $A_B$=7.25 and
$A_V$=5.45 mag for a standard $R_V$=3.1 reddening law.  The large
H$\alpha$/H$\beta$$\sim$60 flux ratio observed before the onset of the dust
obscuration agrees with the large reddening affecting this nova.

\subsection{Distance}

Both the $t_2$ and $t_3$ decline rates and the observed magnitude 15 days
past optical maximum are popular means to estimating the distance to novae.  

The relation between absolute magnitude and $t_2$, $t_3$ takes the form
$M_{\rm max}\,=\,\alpha_n\,\log\, t_n \, + \, \beta_n$.  There are several
different calibrations of $\alpha_n$ and $\beta_n$ available.  The relations
of Cohen (1988) and Downes and Duerbeck (2000, hereafter DD00) for
$t^{V}_{2}$ provide distances to Nova Cep 2013 of 5.1 and 6.3 kpc, and those
of Schmidt (1957) and DD00 for $t^{V}_{3}$ distances of 4.6 and 5.6 kpc,
respectively.  Approximating $m_{pg}$$\approx$$B$, the relation by
Capaccioli et al.  (1989) for $t^{B}_{2}$ gives a distance of 4.7 kpc, and
those of Pfau (1976) and de Vaucouleurs (1978) for $t^{B}_{3}$ result in
distances of 5.5 and 4.7 kpc, respectively.  The average over all above
estimates provides a distance of 5.2 kpc to Nova Cep 2013, while the average
limited to the two determinations from the most recent calibration by DD00
is 6.0 kpc.

Buscombe and de Vaucouleurs (1955) suggested that all novae have the same
absolute magnitude 15 days after maximum light.  The distance to Nova Cep
2013 turns out to be 3.7, 4.4, 3.7, 4.0, 4.6, 4.9, 4.2 and 4.7 kpc using the
calibrations by Buscombe and de Vaucouleurs (1955), Cohen (1985), van den
Bergh and Younger (1987), van den Bergh (1988), Capaccioli et al.  (1989),
Schmidt (1957), de Vaucouleurs (1978), and Pfau (1976), respectively.  Their
average is 4.3 kpc, a distance appreciably shorter than above inferred from
$t_2$ and $t_3$.  This difference is appreciably reduced if the most recent
calibration by DD00 is used for the absolute magnitude 15 days past maximum,
that provides 5.4 kpc for Nova Cep 2013

Another classical MMRD relation uses a specific stretched $S-$ shaped curve,
which is apparent in samples of extragalactic novae.  It was first suggested
in analytic form by Capaccioli et al.  (1989).  Its revision by della Valle
and Livio (1995) provides a distance of 6.4 kpc to Nova Cep 2013, that
increases to 7.0 kpc for the revision proposed by DD00.

Summing up, the distance derived from the magnitude 15 days past maximum
falls appreciably shorter than the minimum distance of $\sim$6 kpc inferred
by the interstellar absorption lines associated to the line-of-sight
crossing the Outer Arm.  Similarly short turns out the distance inferred
from old calibrations of the MMRD based on $t_2$ and $t_3$ times, while the
most recent one from DD00 agrees with the $\sim$6 kpc lower limit.  Good
agreement is also found for distance inferred from the stretched $S-$ shaped
curve for the MMRD relation.  The straight average of the valid distances is
6.5 kpc, that we adopt in this paper and that places Nova Cep 2013 within or
immediately behind the Outer spiral arm.

\subsection{Progenitor}

At that time of the last photometric observation of Nova Cep 2013 logged in
Table~4, +330 days past optical maximum, the optical spectra were totally
dominated by nebular lines (see sect.~6 and Figure~6).  The nebular lines
disappear when the ejecta completely dissolve into the interstellar
space.  If those line are removed from the spectra, the $V$ flux declines by
2.2 mag.  It therefore may be concluded that 330 days past maximum the
central binary was fainter than $V$$>$20.6, in agreement with the absence of
a nova progenitor recorded on Palomar I and II plates.

There is no 2MASS source at the position of Nova Cep 2013. Examining the
2MASS catalog for the 300 sources within 3 arcmin of the nova position, the
detection of infrared sources appear complete to $J$=16.25 and to $K$=15.50,
while the faintest sources score $J$=17.1 and $K$=15.9.  Adopting intrinsic
colors from Koornneef (1983), absolute magnitudes from Sowell et al.  (2007)
and color-dependent reddening relation from Fiorucci and Munari (2003), at
the distance and reddening above estimated for Nova Cep 2013, these limits
completely rule out a luminosity class III giant as the donor star, because
such giant would shine several magnitudes above the 2MASS completeness
threshold.  A K0IV, that represents the faint limit of the subgiant branch,
would shine at $J$=17.0 and $K$=15.4, while a K3IV/III star, that marks the
bright end of the subgiant branch, would shine at $J$=15.4 and $K$=13.8. 
Comparing with the 2MASS completness limit we conclude that the donor star
in Nova Cep 2013 is on or still close to the main sequence.

\section{Spectral evolution}

The spectral evolution of Nova Cep 2013 over the wavelength range of
Cousins' $R_{\rm C}$ and $I_{\rm C}$ bands (5500$-$9000 \AA) is illustrated
in Figure~7, while Figure~6 shows the spectral appearance at bluer
wavelengths at four key times: (a) maximum brightness, (b) close to $t_3$
and before the onset of dust formation, (c) during dust formation, and (d)
one year past optical maximum, when the dust was fully dissolved and the
nova was very faint and returning into the obscurity from which it arose.

The appearance and evolution is that of a typical {\em FeII}-nova. At
maximum, in addition to Balmer, OI, NaI, the most prominent emission lines
were those of FeII, in particoular the multiplets 42, 48, 49, 55, 73 and 74. 
All these lines had deep P-Cyg absorption components at that time
(Figure~6).  With respect to the emission component, on day +1.2 the core
and terminal velocity of the absorption components were $-$675 and $-$1360
km/s for NaI, $-$620 and $-$1270 for OI 7774 \AA, $-$920 and $-$1830 for OI
8446 \AA, $-$800 and $-$1750 for FeII 42, with an average FWHM=800 km/s for
their emission component.  As usual, these velocities grew with time.  On
day +27, the core of the absorption for OI 7774 and 8446 \AA\ reached
$-$2100 km/s, and $-$2500 for NaI.  The evolution of H$\alpha$ and its
absorption systems is illustrated in Figure~9 and discussed in sect.  6.2.

Figure~7 nicely depicts the interplay in the evolution of key emission lines
in the far red spectra of FeII novae.  Close to maximum CaII triplet is in
strong emission, stronger than OI 8446 and with the ratio OI 7774 / OI 8446
$>$1 (in a pure recombination, optically thin case this ratio should be
$\sim$5/3).  At later epochs, OI 8446 first equals and then rapidly
surpasses in intensity the CaII triplet (as a consequence of increasing
ionization of the ejecta), while in parallel the OI 7774 / OI 8446 ratio
rapidly diminishes, an indication of emerging Ly-$\beta$ flourescence.

\subsection{Evolution of emission line flux with time and across dust formation}

Figure~8 highlights the evolution of integrated absolute fluxes (thus
independent of the behaviour of underlying continuum) of representative
emission lines.  For compactness and an easier comparison with the
photometric lightcurves of Figure~4 and 5, the integrated flux of the
emission lines is expressed in magnitudes with respect to the highest
measured value (which is 1.18$\times$10$^{-11}$ erg\,cm$^{-2}$\,sec$^{-1}$
for CaII triplet lines, 1.07$\times$10$^{-11}$ for OI 7774 \AA,
6.42$\times$10$^{-13}$ for FeII multiplet 74 lines, 6.84$\times$10$^{-13}$
for [OI] 6300 \AA, 4.42$\times$10$^{-11}$ or H$\alpha$ and
2.21$\times$10$^{-11}$ for OI 8446 \AA).  Figure~8 is particoularly
interesting because it includes data from the spectrum obtained on day
+40.2, when the dust was already causing an extinction of $\Delta V$=3.1 mag
of the underlying continuum.  In Figure~8, the vertical dashed line marks
the time when dust begun condensing in the ejecta.  The two curves, one
solid and one dotted, that repeats identical in each panel, show for the $V$
band the observed and expected decline in absence of dust formation,
respectively (imported from Figure 5).  These two curves have been
scaled to the position expected for a given emission line at the time of the
start of dust condensation, obtained from a low order polynomial fit to the
earlier data.  There is of course some arbitrarity in this extrapolation and
an actual spectrum taken closer in time to the start of dust condensation
would have been most useful in this regard, but unfortunately none is
available. The aim of Figure~8 is to investigate the relative location 
of emission lines, continuum and dust formation sites within the
ejecta assuming a spherically symmetric shape for them. Similar plots of
the evolution with time of the integrated fluxes of selected emission lines
is presented by Munari et al.  (2008) for the slightly faster FeII nova
V2362 Nova Cyg 2006 which also formed dust, but in much lower quantity than
for Nova Cep 2013.

  \begin{figure}
    \centering   
    \includegraphics[width=8.5cm]{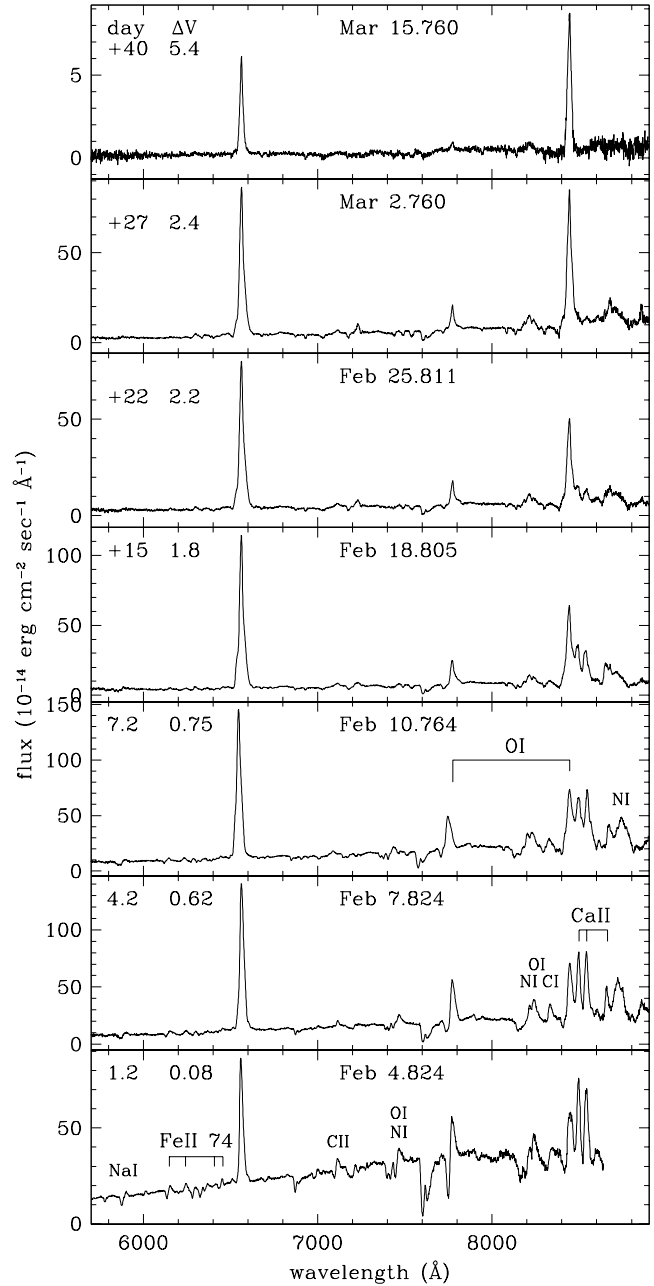}
     \caption{Spectral evolution of V809 Nova Cep 2013 at red wavelegths.
     $\Delta V$ lists the decline inmagnitudes from maximum brightness.}
     \label{fig1}
  \end{figure}  

For the following discussion, it is relevant to note that the degree of
ionization of the ejecta was not appreciably changing across the onset of
dust condensation.  In fact, In Figure~6 on the spectrum for day +40.2 no
significant [OIII] emission lines are seen.  Appearance of strong [OIII]
lines normally mark the transition from optically thick to optically thin
conditions (McL60), when high ionization rapidly spreads through the ejecta. 
Shore and Gehrz (2004) suggested that incresing ionization induces rapid
grain growth in novae, but this did not happened in Nova Cep 2013.  The
+40.2 spectrum was obtained about four days past onset of grain condensation,
when the dust was already causing a large reduction in the photo-ionizing
input flux from the underlying pseudo-photosphere.  Were four days enough to
significantly affect the ionization balance in the ejecta ?  Actually not. 
In fact, the recombination time scale of the ionized ejecta goes like
(Ferland 2003):
\begin{equation}
t_{\rm rec} = 0.66 \left(\frac{T_{\rm e}}{10^4 {\rm ~K}}\right)^{0.8}
\left(\frac{n_{\rm e}}{10^9 {\rm ~cm}^{-3}}\right)^{-1}  ~~~~{\rm (hours)}
\end{equation}
where $T_{\rm e}$ and $n_{\rm e}$ are the electron temperature and
density, respectively.  The presence of a strong [NII] 5755 emission line in
the day +40.2 spectrum suggests that the electronic density was below the
critical value for this line at 6$\times$10$^{4}$ cm$^{-3}$, which
corresponds to a recombination time scale of 1 year for any reasonable
assumption about the electronic temperature.  In summary, from the point of
view of the photo-ionization balance, the ejecta evolved smoothly between
the two spectra obtained on day +29.2 and +40.2.

The behaviour of CaII triplet and OI 7774 \AA\ represents the extrema
of the observed evolution of emission lines across the dust formation
episode.

  \begin{figure}
    \centering   
	    \includegraphics[width=8.5cm]{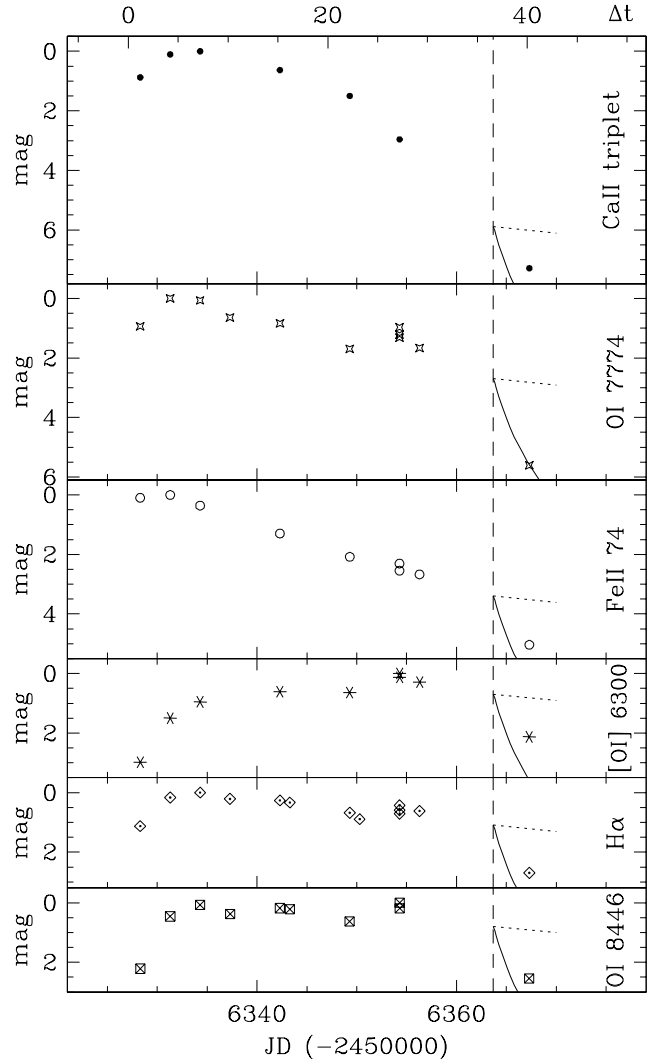}
     \caption{Evolution of the integrated flux of some representative emission lines
      of V809 Nova Cep 2013. The size of the error bars is similar to symbol
      dimension.}
     \label{fig1}
  \end{figure}  

CaII triplet lines reached their maximum flux $\sim$7 days past optical
maximum.  Their evolution was very smooth, essentially unaffected by the
condensation of dust.  CaII is easily ionized and the triplet lines form
from low excitation levels, $\sim$3.13 eV above ground state.  As ionization
progresses through the ejecta along the decline from maximum, the location
of formation of CaII lines is therefore rapidly pushed outward.  Figure~8
tells us that the dust condensed in the part of the ejecta {\em internal} to
the outer layer from where CaII lines originate.

The flux of OI 7774 \AA\ line dropped by as much as the continuum radiation,
indicating that it formed in a region of the ejecta close to the
pseudo-photosphere and fully internal to the layer where dust formed.  OI
7774 \AA\ forms from a high upper energy level, 10.7 eV above ground state,
populated mainly during recombination of OII.  The proximity to the
pseudo-photosphere is confirmed by the continued presence of P-Cyg
absorption to this line even on spectra for both day +29.2 and +40.2, when
P-Cyg absorptions were already gone for all the other lines and still weakly
present only in Balmer series lines.

The behavior of these two lines confines the region of dust formation to the
central layer of the ejecta.  The attenuation across dust condensation of
the other lines considered in Figure~8 (between 1.0 and 1.5 mag), indicates
that the region where they formed is the same where dust formed.  In fact,
assuming that in the region of dust formation the density of dust and of
emitting gas scales similarly with distance, an extinction of 1.45 mag is
expected for a line forming exactly in the same layer as the dust, of 1.3
mag if 10\% of the line flux forms above the dust layer, of 1.1 mag if this
proportion rises to 20\%.

\subsection{Evolution of H$\alpha$ profile and its absorption systems}

The evolution of the H$\alpha$ emission line profile before the onset of
dust formation is shown in Figure~9 from Echelle high resolution
observations.  During the first week, the profile is simple and
characterized by the usual combination of a broad emission component and
various blue-shifted and sharper superimposed absorption components
(Payne-Gaposchkin 1957, McL60, Munari 2014).  Later on, an unusual central
peak appeared superimposed to these standard components, and this peak
became progressively sharper, from FWHM=500 km/s on day +10.2 to 210 km/s on
day +27.2.  The broad emission and the multiple absorption components of the
H$\alpha$ profiles have been fitted with Gaussians and the resulting
fit superimposed in Figure~9 to the actual observed profile.  The fit is
particularly good and the heliocentric velocity and width of the individual
Gaussians are listed in Table~5.

The radial velocity of the emission component became less negative and its
width increased with time.  This is the effect of a progressively decreasing
optical thickness of the expanding ejecta, that allows more direct radiation
from internal strata and from the receding side to reach the observer.

On the first spectrum of Figure~9, two absorption components are present. 
They correspond to the {\em principal} absorption system described by McL60
from old photographic spectra and by Munari (2014) from modern CCD
observations.  This absorption system is normally observed as a single
component, but sometimes - as in this case - is splinted in two.  Other
novae with splinted {\em principal} absorption lines were, for ex.,  Nova
Gem 1912, Nova Aql 1918 and Nova Cyg 2006.  The second spectrum in Figure~9
was obtained at day +4.2, and shows the appearance at larger velocities of
the {\em diffuse-enhanced} absorption system, which is characterized by a
single component at all epochs covered by our observations.  As the
ionization spreads through the ejecta, the neutral regions where these
absorptions can form have to move outward in mass, involving progressively
faster moving ejecta, with the result that the velocity of the absorptions
grows more negative with time.  This is well confirmed by Figure~10 that
plots versus time the velocities of the various absorption components listed
in Table~5, and shows how they follow nice linear trends.  Similar linear
trends have been reported for other recent FeII dust forming
novae, like Nova Cyg 2006 (Munari et al.  2008) and Nova Scuti 2009 (Raj et al. 
2012).

  \begin{figure}
    \centering   
    \includegraphics[width=8.5cm]{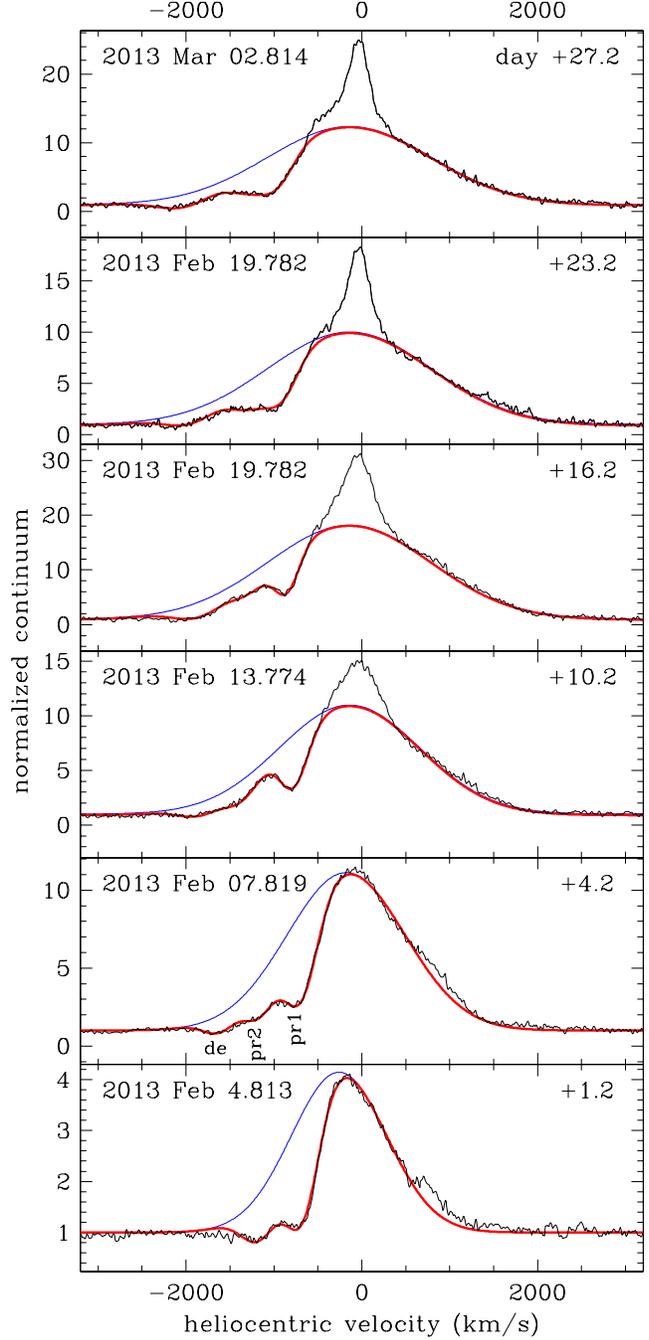}
     \caption{Evolution of the H$\alpha$ profile of V809 Nova Cep 2013.
     The multi-component fit, listed in Table~5 and described in sect 6.2, is
     overplotted in red. The blue line is the Gaussian fitting to the broad
     emission component, truncated by absorption components at shorter
     wavelengths (identified by {\em pr1}, {\em pr2} and {\em de} in the panel
     for 2013 Feb 07.819).}
     \label{fig1}
  \end{figure}  
  \begin{table}
    \centering      
     \caption{Heliocentric radial velocity and $\sigma$ (km/sec) of the Gaussian fitting 
     to the absorption components of the H$\alpha$ profiles of V809 Nova Cep 2013 shown 
     in red in Figure~9.}
    \includegraphics[width=8.0cm]{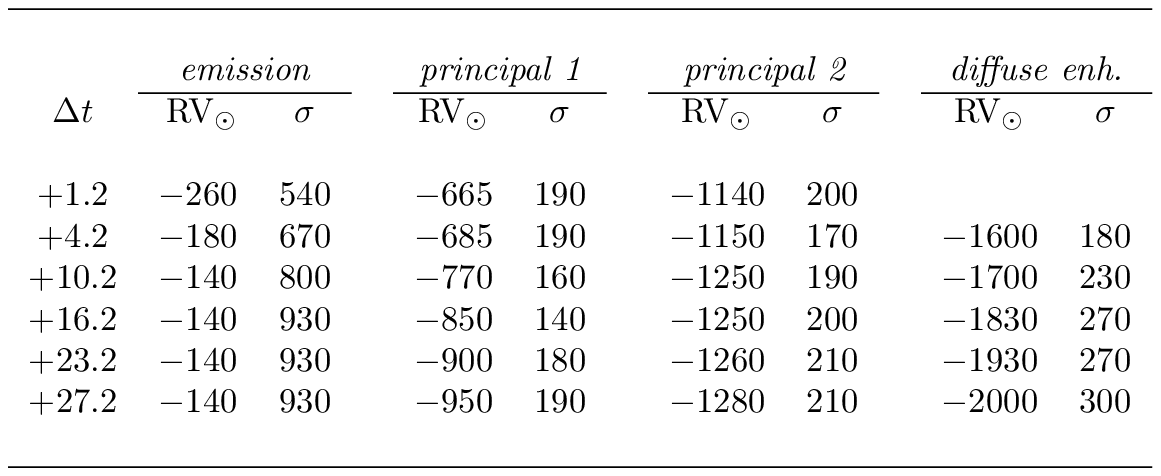}
     \label{tab2}
  \end{table}    
  \begin{figure}
    \centering   
    \includegraphics[width=6.5cm]{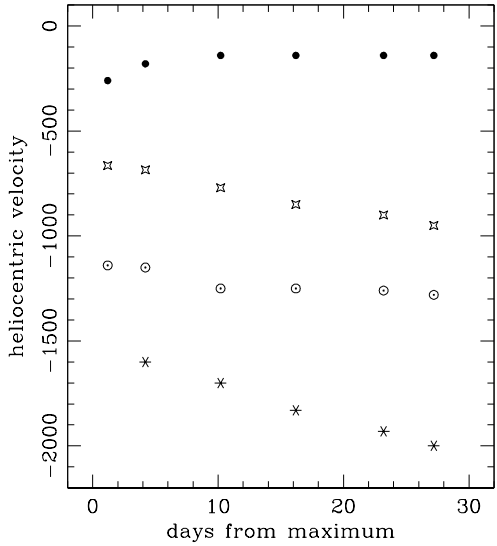}
     \caption{Time evolution of the heliocentric radial velocity of the
     H$\alpha$ components, listed in Table~5 and fitted to the line profiles
     in Figure~9.  From top to bottom the symbols indicate: emission
     component, principal P1, principal P2 and diffuse enhanced absorption
     components.}
     \label{fig1}
  \end{figure}  

Data summarized by McL60 show a correlation between the {\em mean} radial
velocity of the various absorption systems and the speed class of the nova. 
The McL60 velocity relation for the principal system is $\log v_{\rm
prin}$ = 3.57 $-$ 0.5$\log t_2$, and predicts $\approx$$-$930~km/s for Nova
Cep 2013, in excellent agreement with the observed $-$960~km/s, which is
obtained by averaging components 1 and 2 in Table~5.  The
McL60 relation for the diffuse enhanced system is $\log v_{\rm
dif-enh}$ = 3.71 $-$ 0.4$\log t_2$, and predicts $\approx$$-$1700~km/s
velocity for Nova Cep 2013, in good agreement with the $-$1800~km/s average
of the value reported in Table 5.  It is worth noticing that the {\em
pre-maximum} absorption system, clearly present in early epochs spectra
of Nova Cyg 2006 and Nova Scuti 2009, is apparently missing in Nova Cep
2013.  The McL60 $v_{\rm pre-max}$$\approx$$-$4750/$t_2$ relation indicates,
in the case of Nova Cep 2013, a velocity of $-$250 km/sec for the
pre-maximum system. This is close to the velocity of the narrow component
that became obvious at later epochs, so there could be some
filling-in at earlier epochs that prevented detection of the pre-maximum
absorption system.

  \begin{figure}
    \centering   
    \includegraphics[width=8.0cm]{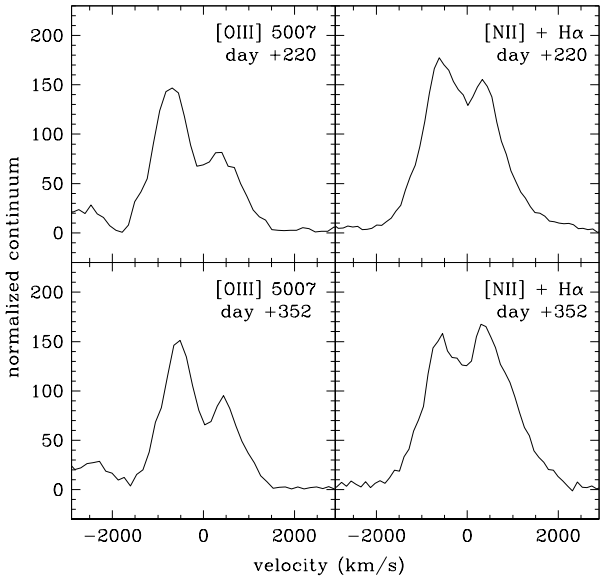}
     \caption{Line profiles for [OIII] 5007 \AA\ 
      and the blend [NII] 6548, 6584 + H$\alpha$ from low resolution
      spectra for 2013 Sep 11 (day +220) and 2013 Dec 30 (day +330), 
      after Nova Cep 2013 had reemerged from dust obscuration.}
     \label{fig1}
  \end{figure}  

When Nova Cep 2013 emerged from dust obscuration, the spectrum had turned
into a deep nebular one, and was dominated by [OIII] 4959 and 5007, [NII]
5755, 6548 and 6854, and [OII] 7325 \AA\ (cf. Figure~6), with only a
feeble trace of H$\beta$ still visible. Figure~11 presents the
emission line profiles (from low res spectra) for [OIII] 5007 \AA\ and [NII]
6548, 6584 \AA\ + H$\alpha$ at days +220 and +352, now dominated by a double
peak with a velocity separation of 1160 km/sec. This traces an outward bulk
velocity of just $\sim$600 km/s, largely lower than that characterizing the
absorption systems and the broad emission seen in the early phases, reaching
$\sim$2000 and $\sim$1000 km/sec, respectively. 

This is a manifestation of the dilution with time of the ejecta into the
interstellar space.  At early phases, the emission line profiles are
dominated by the outer and faster moving material, and radiation from the
inner part of the ejecta does not reach the observer because of the large
optical depth of outer ejecta.  As the ejecta expand and thin, more
radiation from the inner and slower regions of the ejecta contribute to the
spectra.  The emissivity of the gas is proportional to the number of
recombinations per unit time, i.e.  to the local electron density that goes
down like $r^{-3}$ ($\propto$ $t^{-3}$) for an initial ballistic launch of the
ejecta.  Consequently, at the time of spectra at days +220 and +352, the
emissivity of the outer regions was essentially nulled by the great
dilution, and only the inner and slower ejecta were still able to contribute to
observed spectra, resulting in the much slower expansion velocity inferred
from the separation of the double peaked emission lines of Figure~11.

\subsection{Late appearance of a narrow component in H$\alpha$}

One unusual spectroscopic feature of Nova Cep 2013 deserves to be
highlighted: the appearance of a narrow component superimposed on the broad
underlying profile of H$\alpha$.  As shown in Figure~9, this component was
first visible at FWHM=500 km/sec on the H$\alpha$ profile for day 10.2.  The
component rapidly narrowed, reaching FWHM=210 km/sec on day +27.2.  The low
S/N of the other emission lines on our Echelle spectra limits the
possibility to assess the presence of the narrow component on other lines. 
We can only say that it was surely present for the OI 8446 \AA\ line, and
perhaps also for the OI 7774 \AA\ line.  The detection of the narrow
component on the H$\alpha$ profile has been possible only thanks to the high
resolution of our Echelle spectra.  Lower resolution data - as normally
obtained for novae - would have missed it entirely.

Attention to narrow components is a recent affair. Their presence in Nova
Oph 2009 was modelled by Munari et al.  (2011) with an equatorial ring in a
prolate system with polar blobs, in Nova Mon 2012 by Ribeiro et al.  (2013)
as a density enhancement toward the wrist of a bipolar structure.  In some
other novae the sharp component is instead believed to originate from the
central binary and to trace (restored) accretion (Mason et al.  2012, Walter
\& Battisti 2011).  Nova KT Eri 2009 seems to be a transitional case,
showing the narrow component coming from both the ejecta as, at later times,
from the accreting central binary (Munari, Mason and Valisa 2014).

The narrow component in Nova Cep 2013 become progressively visible as the
outer ejecta were thinning and the inner ejecta were emerging into view.  It
originates from spatially structured inner ejecta, and cannot be associated
with emission from the central binary for two basic reasons: (1) it was
observed at a time when the ejecta were still optically thick and thus
blocking direct view of the core of the system.  The large optical thickness
at that time is confirmed by the non detection of super-soft X-ray radiation
(Krautter 2008) from the nova in the Swift observations by Chomiuk et al. 
(2013) for Feb 22.1 (day +18.5), when the narrow component was already
strong, and (2) the flux radiated by the narrow component amounts to
3.7$\times$10$^{-12}$ erg\,cm$^{-2}$\,sec$^{-1}$ (straight average over the
similar values for day +10.2, +16.2, +23.2 and +27.2).  Transforming this
flux into a photometric magnitude, the narrow component alone would shine as
a star of $R_{\rm C}$$\sim$15.0 mag.  Considering that the progenitor was
fainter than the $\sim$20 mag plate limit for Palomar II red plates, the
narrow component alone radiated in the $R_{\rm C}$ band $>$100$\times$
more than the progenitor in quiescence, a condition incompatible with an
origin of the narrow component from the central binary.

\section{Acknowledgements}
We are grateful to Elena Mason for useful discussions and careful reading
of the manuscript.

\setcounter{table}{3}
  \begin{table*}
    \centering      
     \caption{Our $B$$V$$R_{\rm C}$$I_{\rm C}$ of V809 Nova Cep 2013.}
    \includegraphics[width=14cm]{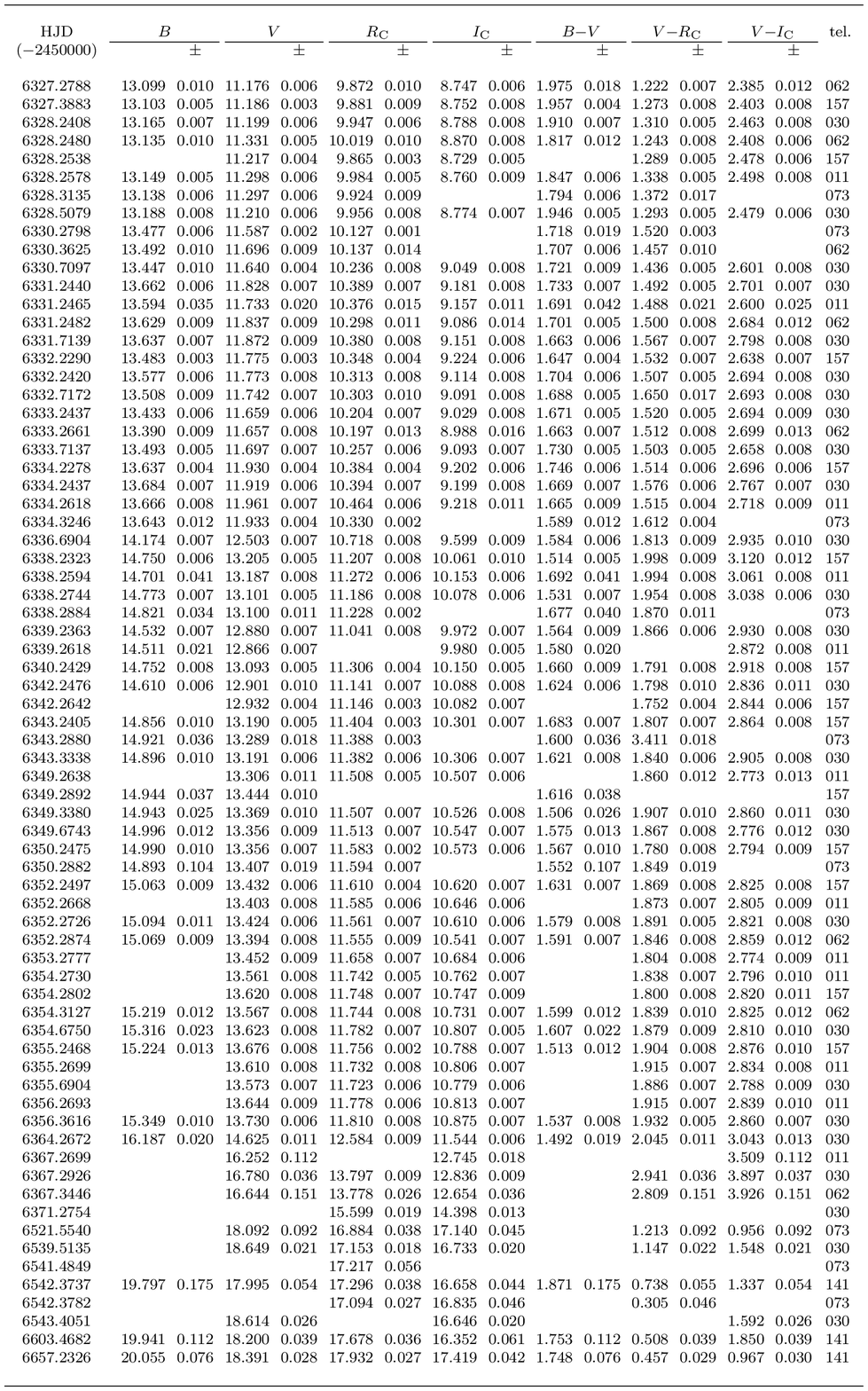}
     \label{tab2}
  \end{table*}    

\bsp

\label{lastpage}

\end{document}